\documentstyle[preprint,aps]{revtex}
\input{epsf}
\draft

\newcommand{\be}{\begin{equation}}
\newcommand{\ee}{\end{equation}}

\newcommand{\fig}[1]{Fig.~\ref{#1}}
\begin{document}
\title{Mesoscopic Analysis of Structure and Strength of Dislocation 
Junctions in FCC Metals}
\author{V.~B.~Shenoy$^1$, R.~V.~Kukta$^2$ and R.~Phillips$^1$\\
{\small $^1$Division of Engineering, Brown University, Providence, RI
02912}\\
{\small $^2$ Division of Eng. and
Applied
Science, California Institute of Technology, Pasadena, CA 91125}\\
}
\date{\today}
\maketitle

\begin{abstract}
 We develop a finite element based dislocation dynamics model
to  simulate
 the structure and strength of dislocation
junctions in FCC crystals. The model is based on  anisotropic
elasticity theory supplemented by the explicit
inclusion of the separation of perfect dislocations into 
 partial dislocations bounding a stacking fault. We demonstrate that the  model
reproduces in precise detail the structure of
the Lomer-Cottrell lock already obtained from atomistic simulations. 
In light of this success, we also examine the
strength of junctions culminating in a stress-strength diagram which is the
locus of points in stress space corresponding to dissolution of
the junction.
\end{abstract}


\vspace{.5cm}
 In FCC metals, a key mechanism limiting the
movement of the dislocations is the ``forest intersection''
mechanism, where  segments of dislocations on a glide plane are rendered
 immobile as a result of intersection with dislocations on other glide
planes. Such intersections can lead to complex
{\it dislocation junction} structures since the cores of the
dislocations in
these metals are
dissociated into partial dislocations separated by a stacking
fault\cite{hirthlothe}. Furthermore, the structure of the
junction also depends strongly on the geometric disposition (such as the
line directions and Burgers vector) of the participating
dislocations. A core level analysis of all the possible junction
configurations is therefore important from a number of perspectives. Such
core level calculations in conjunction with statistical averaging procedures
can provide key parameters that are used in models that predict the 
mechanical behavior of metallic crystals on a macro-scale. 
These models include
single crystal plasticity
models\cite{bassani} 
and the computational models that simulate the dynamics of a large
collection of dislocations\cite{fivel}.    
In this letter we develop a mesoscopic dislocation dynamics model that
can be used to simulate 
the structures and strength of  dislocation junctions in FCC metals
and may obviate the need for direct atomistic simulations of
these junctions.

For  simple dislocation intersection geometries,
the structure of  dislocation junctions has been studied extensively
in a series of classic papers by Hirth and
coworkers\cite{hirthco} using the theory of linear elasticity.
Analytical insights into more complicated intersection geometries
have been gained by using the line tension approximation for the 
dislocation lines\cite{saada}. While this approach provides
a great deal of physical insight into the junction structure
and strength, it  ignores the extended core structure
of the dislocations as well as the long range interaction between the
dislocation segments. With rapid advances in computational power in
recent years, it has become possible to perform atomistic
simulations of dislocation intersections\cite{bulatov,zhou,rodney}.
Typically these simulations\cite{zhou,rodney}, performed using cells of the
dimensions
of 300-400$\AA$ containing over a million atoms, are computationally
very demanding and raise serious questions concerning the
role that boundary conditions play in dictating the results. 

In the present study, we develop a mesoscopic method
to study the structure and strength of  dislocation junctions
  that includes the dissociation of the dislocation
core into partial dislocations. The interaction between the dislocations are treated using the
theory of anisotropic elasticity\cite{bacon}. We find that the our method 
 reproduces, in precise detail, all the features of the dislocation
junction structure obtained from a full atomistic treatment of the
dislocation core. Our results demonstrate that
 the junction structure is almost entirely 
determined by elastic interaction between the partial dislocations
 and the stacking fault energy. Furthermore,
the adaptive procedures that we use allow us to simulate dislocation
junctions that have lengths of the order of microns.
 An atomistic simulation of these large junctions
would require simulations involving an excess of a billion atoms. 
 In order to clearly demonstrate the role of the stacking fault in
determining the junction structure, we will
consider dislocation junctions in two metals, namely, Al, with a high stacking fault energy
(0.104 $J/m^2$) and
Ag, with a low stacking fault energy(0.016 $J/m^2$)\cite{stack}. 
We limit our
discussions to 
the Lomer-Cottrell lock\cite{hirthlothe}; a complete investigation of
other junctions will be reported elsewhere.

The simulations are carried out through an adaptive finite
element based dislocation dynamics algorithm that is described in detail in
Ref.\cite{kukta}. We start our simulations with
 straight dislocation lines that are pinned at their end
points. Simulations
are carried out  until the dislocations glide to
their equilibrium configuration. Each dislocation line is allowed to split into
partial dislocations by accounting for the energy cost of the
stacking fault created in the process.
In the starting configuration, the partial dislocations
are discretised into straight segments of equal
length. A time step of the dynamics process consists of moving the node
connecting the segments with a velocity that is proportional to the nodal
driving force. The computation of the nodal force requires the
knowledge of the force per unit length at certain quadrature
points on the dislocation segments
that are attached to the node. The force per unit length at any point
on a segment consists of a component arising from the stresses due
to all the dislocation segments in the system including the segment
itself. The Brown regularization procedure is adopted to guarantee
that the self-stress contribution is well
behaved\cite{schwarz}. When the interaction between the segments
belonging to different dislocations is attractive, they 
 approach  each other in the process of
forming a junction. The stress acting on one of the segments
due to the other one
diverges as their separation vanishes. 
In our calculations,
 the stresses are constant for distances less than a critical
separation distance, $r_c$,  by the value of stress computed at 
$r_c$. 
As a result, the segments are locked once they are closer
than this critical distance; a junction has formed. 
From an elastic perspective, 
at distances larger
than $r_c$,  the stresses produced by the
junction segments correspond to those produced by a dislocation whose Burgers
vector is the sum of the Burgers vectors of the two segments that
make up the junction. Once the junction is formed, it can unzip if the
external stresses
cause the segments that form the junction to move away from each other.
The calculations described here were carried out with $r_c=b=a/\sqrt{2}$,
although we have also considered the cases in which the cutoff was
$b/2$ and $2b$, without noticeable change to the resulting junction
structures.
In addition to the stress from the dislocation
segments, the force per unit length consists of a component arising from
 the stacking fault. This component
 is normal to the dislocation segment and has a magnitude that
equals the stacking fault energy and 
acts in a direction tending to shrink the stacking fault. In the
simulations that we describe, we have ignored the frictional stress since
we have found that its inclusion is unimportant.
 A  key feature of our simulations is the
adaptive positioning of the nodes. As the simulation progress, the
nodes are redistributed, so that
the  regions with large curvature have more nodes per unit length. 

\begin{figure}
\centerline{\epsfysize=7truein \epsfbox{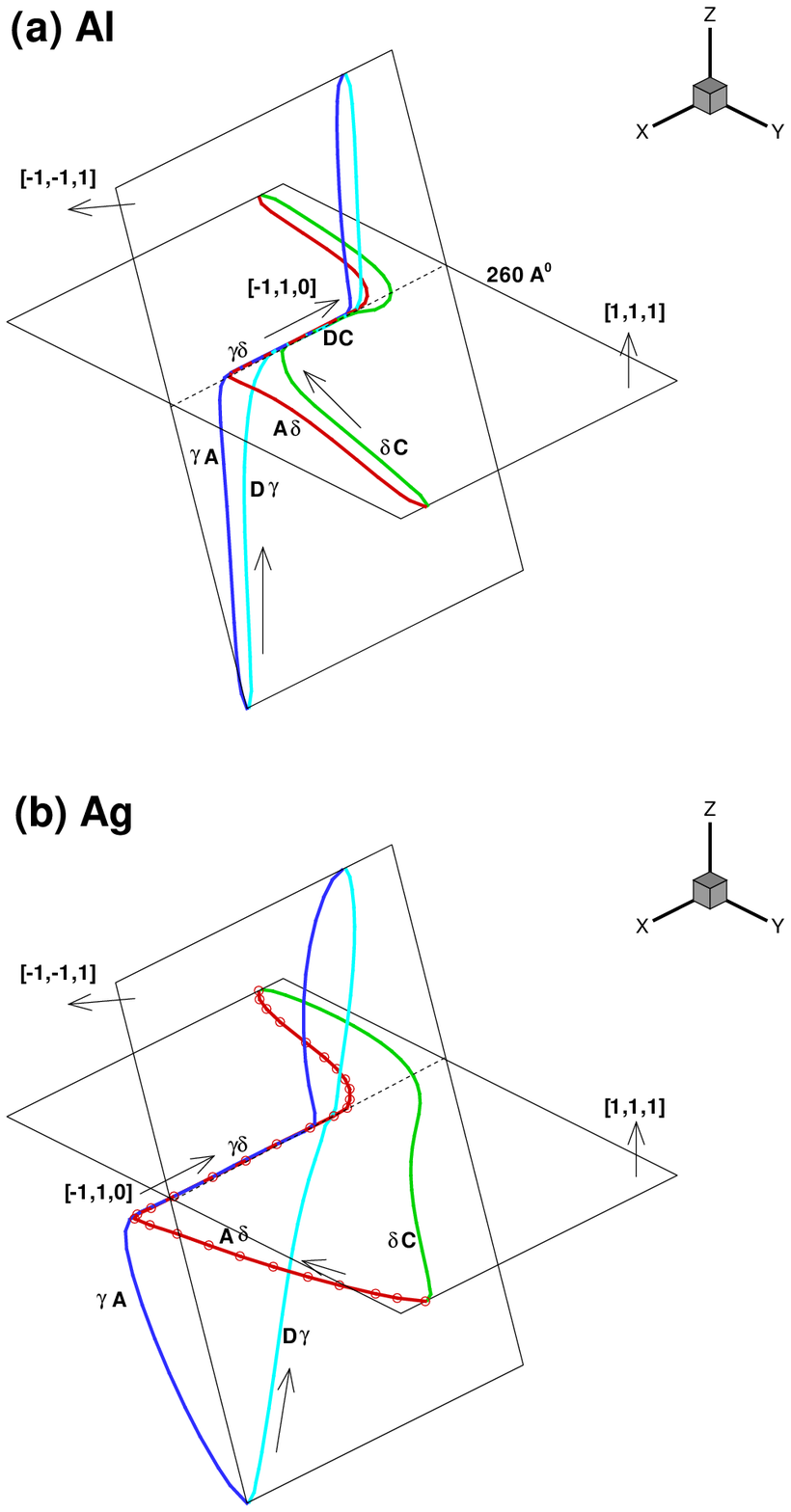}}
\vspace{-3.5cm}
\caption{Structure of the Lomer-Cottrell Junction in Al and Ag. The
line directions and the Burgers
vectors of the dislocation segments are indicated in the figure. The
junction forms along the dotted line, which is the line of
intersection of the two slip planes.}
\label{lc1}
\end{figure}

We demonstrate our results by first considering the  
equilibrium configuration of a Lomer-Cottrell lock as shown in \fig{lc1}.
 This configuration has been chosen so as to
make a direct comparison of our results with the atomistic simulations
for Al reported in Ref.\cite{rodney}. The pinning points
are arranged such that in the starting configuration 
the dislocation line directions  make an  angle of $60^0$ with
the $[\bar{1},1,0]$ direction. This direction coincides with
the line of intersection of the  slip planes of the dislocations that
form the junction. The initial
separation between the dislocations was chosen to be 6$\AA$.
 The line directions of each of the partial
dislocations and their
slip plane normals  are given in the figure. We
follow the notation described  in Ref.\cite{hirthlothe} to label the Burgers
vectors of the partial dislocations by referring to the Thompson tetrahedron.
For example, the $a/2[0,1,\bar{1}](111)$ dislocation splits into partial
dislocations $A\delta$ and $\delta C$ with Burgers vectors
$a/6[\bar{1},2,\bar{1}]$ and $a/6[1,1,\bar{2}]$ respectively.
As is evident from the figure, the junction segment in
the case of Al, has split into separate parts. A stair-rod 
 segment with  Burgers vector $\gamma\delta$ of the type
$a/6\langle110\rangle$ forms an extended node on the left side of the junction.
The remaining part of the junction is a sessile Lomer dislocation
segment. The length of the stair-rod segment is 38$\AA$, while the 
Lomer is 42$\AA$, which is in excellent agreement with the 
atomistic results\cite{rodney}.
The dislocation dynamics model also agrees with the 
atomistic results in predicting the structure
of the right hand node, which is point-like and 
is the meeting point of constricted dislocation segments.

As an extension of earlier results and to highlight the dependence
of the junction structure on the stacking fault energy, we have also
computed the geometry of the Lomer-Cottrell junction in Ag.   The dislocation
junction in Ag, for this configuration, has an entirely different structure.
The junction segment is entirely composed of a stair-rod dislocation of
length $180$\AA.  The
smaller stacking fault energy keeps the segments
$\delta C$ and $D\gamma$  from participating in the junction
formation process. Before proceeding to discuss the effect
on junction structure of the length and the orientation of the
participating dislocations, we pause to illustrate
the adaptive nodal repositioning process used in the simulations.
It can be seen from Fig.1b that the nodes on the red
dislocation line  have evolved from a case where they are
spaced at equal intervals to a case where there are more nodes in 
the regions with large curvature.

In order to consider the effect of
varying the distance between pinning points and the orientation of
the dislocations on the junction structure, we consider the 
highly symmetric geometry  shown in \fig{lc1}, 
where the 
participating dislocations have the same orientation and identical
distances between the pinning points. 
In the case of the dislocation junction in Al, we find that  
 the length of the Lomer segment increases with increasing distance
between the pinning points, while the length of the stair-rod segment
is not altered. By way of contrast for the junction in Ag, the length of the
stair-rod keeps increasing on increasing the distance between the
pinning points up to about 1000$\AA$. Beyond this length,
the Lomer segment appears and increases in length, while the
left hand stair-rod node attains a constant length of about 600$\AA$.
We therefore conclude that the
length of the extended node is determined largely by the
stacking fault energy of the crystal. 
This calculation also illustrates the ability of
our method to handle large junction lengths which are clearly 
beyond the reach of the atomistic simulations.

\begin{figure}
\centerline{\epsfysize=3.5truein \epsfbox{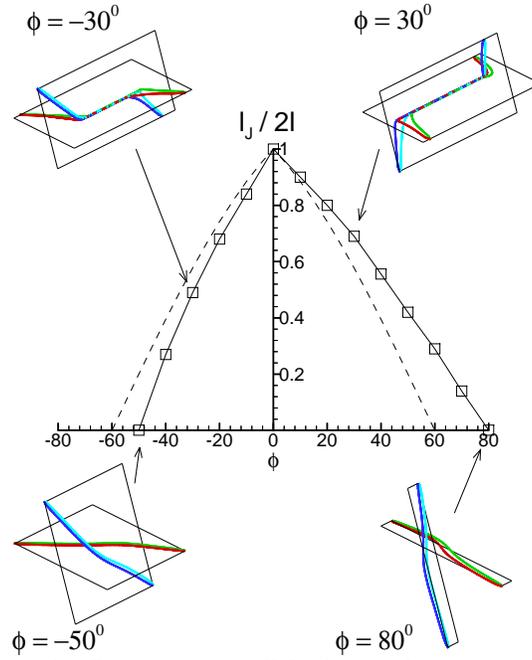}}
\caption{Junction length of the Lomer Cottrell lock in Al
as a function of the line directions of the participating
dislocations. The dotted line is the prediction of the line-tension
model and the ``squares'' are the results obtained using our
mesoscopic model. We also show  the evolution of 
the junction structure 
as the line directions are varied.}
\label{angle}
\end{figure}

 We  now consider the effect of altering the junction angle on
the structure of the dislocation junction. We use $\phi$ to 
denote the
angle  between the dislocation line direction in the
starting configuration 
and the $[\bar{1},1,0]$ direction, $l_{J}$ to represent the junction
length and $2l$ for the distance between the pinning points.
In \fig{angle}, we plot the junction length and
structure  as the orientation of the pinning points is
altered. The dislocation junction ceases to form at  angles smaller
than -50$^0$ or angles greater than 80$^0$. For comparison the 
we also show the results of the line tension model
tension model of Saada\cite{saada}. Within this model the junction length
vanishes for angles that are larger than 60$^0$ or smaller than
-60$^0$. The difference can be understood by noting the difference in
the node structures in junctions with negative angles in \fig{angle}.
The left node for the
-30$^0$  configuration is made up of
a stair-rod of the type $\gamma \delta$,
while the right node is made up of a new stair-rod segment
$\delta D/C\gamma$, which is a $a/3\langle110\rangle$ type dislocation. 
The existence of the $a/3\langle110\rangle$ in the negative angle
junctions is counterintuitive, since
a simple line tension estimate\cite{hirthlothe} shows that
this type of stair rod has 4 times higher energy per unit length than
the $a/6\langle110\rangle$ stair-rod. The dislocation interactions
that are ignored in this simple line tension argument conspire to
make this stair-rod segment stable. The asymmetry in the junction formation
angles can be attributed to this difference in junction structures of
the positive and negative angle junctions.

\begin{figure}
\centerline{\epsfysize=3.5truein \epsfbox{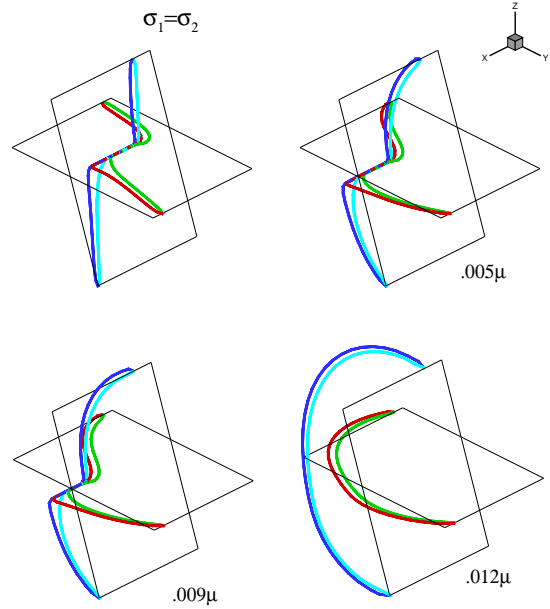}}
\caption{Evolution of the symmetric Lomer Cottrell lock in Fig.1a,
on applying an external stress. The resolved shear stress on the two
junctions are the same.}
\label{stress}
\end{figure}

\begin{figure}
\centerline{\epsfysize=3.0truein \epsfbox{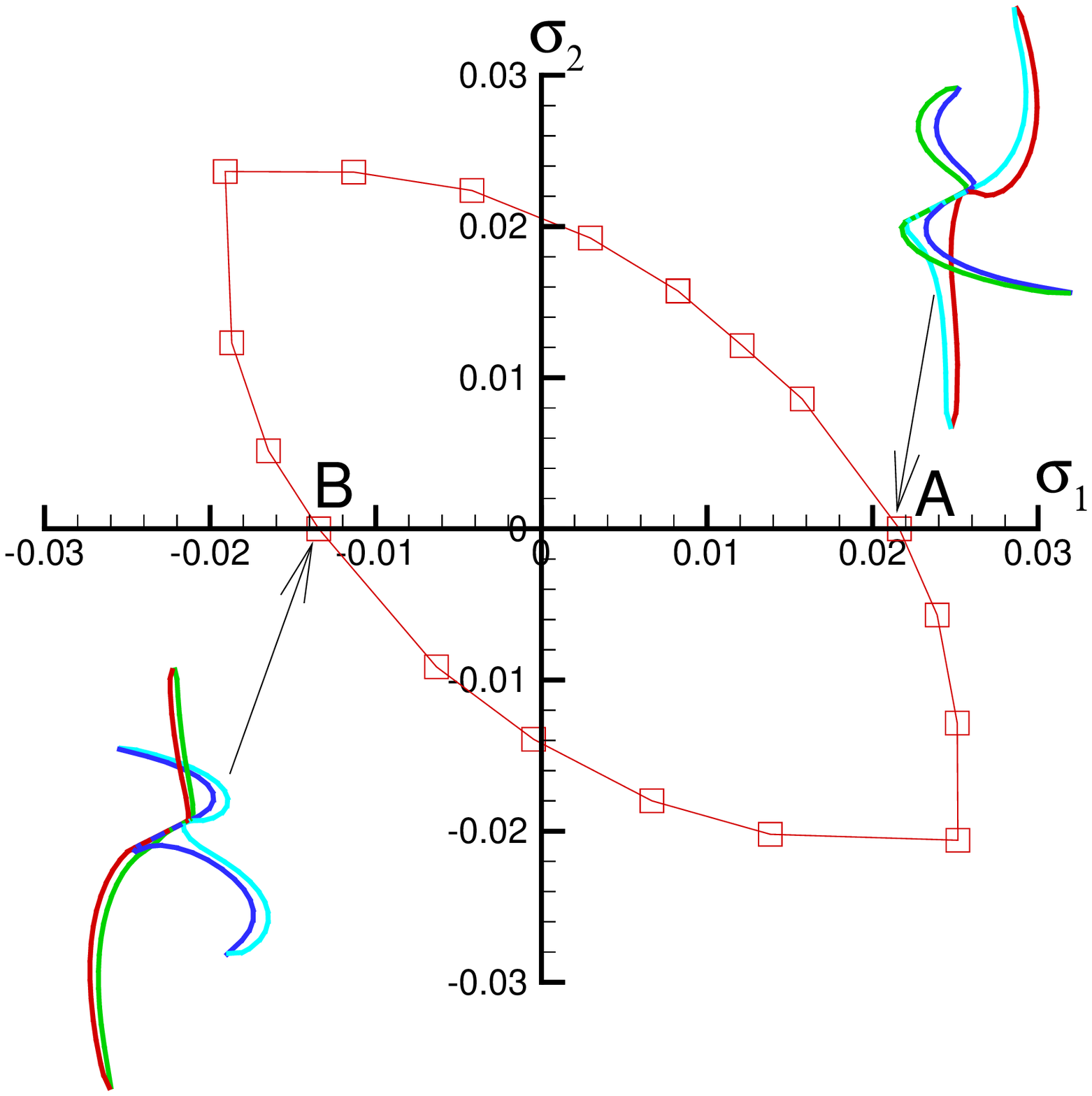}}
\caption{The ``yield surface'' for the symmetric Lomer-Cottrell
lock in Fig.1a. The stresses are in the units of shear modulus of
Al. We also display the structure of the junctions prior to
destruction for the points marked A and B.}
\label{yield}
\end{figure}

 Far more interesting than the structure of junctions 
 is their behavior under stress.   Line tension models predict that   the arms
of the dislocations bow out under the influence of the external stress,
while the junction translates along the line of intersection of the
slip planes. As the junction
translates, its length decreases as a result of the applied stress via
an ``unzipping'' mechanism. \fig{stress} shows the evolution of the
dislocation junction in Fig.1a under the influence
of an externally applied stress. The resolved shear stress on the dislocations in
the $(1,1,1)$ and $(\bar{1},1,1)$ slip planes are labeled $\sigma_1$ and $\sigma_2$
respectively. We have chosen the orientation of the applied stress for this
series of pictures 
such that $\sigma_1 = \sigma_2$. On increasing the stress, the length of the Lomer
segment initially increases on going from no applied stress to a
stress of 0.006$\mu$ ($\mu$ is the shear modulus). This 
is consistent with the atomistic simulations, but is at variance with
the line tension models, which always predict a decreasing junction
length with increasing stress. 
On increasing the stress further, the junction
translates and undergoes an unzipping mechanism, whereby the length of
the Lomer segment decreases. This behavior is evident for the stress level of
.009$\mu$. On increasing the stress further, the junction breaks at
a stress level of about 0.012$\mu$ after which they continue to bow out. The
configuration shown at stress of 0.012$\mu$, is not in equilibrium,
 but a snap shot after the junction has been destroyed. We have
carried out similar calculations for several values of resolved
shear stress acting on the two dislocations. A ``yield surface''
for dislocation destruction is shown in \fig{yield}. This surface is
symmetric about the line $\sigma_1 = \sigma_2$, but depends on the
sign of $\sigma_1$ and $\sigma_2$. This symmetry breaking is readily
understood by looking at the configurations of the dislocations
under action of positive and negative $\sigma_1$(denoted in
\fig{yield} by points A and B respectively) with $\sigma_2=0$.\
The presence of the stair-rod node
makes it more difficult for the horizontal segment to bow out in case
(A) compared to case (B) and
requires  a larger value of  shear stress to break the
junction. 
    
  It is interesting to compare the predictions of the
line tension model and the atomistic simulations
with the results obtained from our simulations.
An important prediction of the line tension model is the scaling of
the breaking stress of symmetric junctions  with the distance between the pinning points,
written as $2l$. The line tension model predicts  the average 
critical resolved shear stress to break the symmetric junctions to be $\approx
0.5 \mu b/l$. We have
simulated the breaking of symmetric junctions  with lengths
ranging from 300$\AA$  to 1$\mu m$ by applying the external stress
in different orientations.  We have found that 
the critical breaking stress scales as $\sigma_c \sim \mu b/l$. For the cases
$\sigma_1 = \sigma_2$ and $\sigma_1 = -\sigma_2$ the critical
resolved shear stress behaves like $\sigma_c = 0.64 \mu b/l$
and $\sigma_c = 1.24 \mu b/l$, respectively. For the
$300 \AA$ junction, 
for the case when $\sigma_1 = 1.3 \sigma_2$, the atomistic
simulations  gave 
$\sigma_c = .017\mu = 0.8\mu b/l$, which is in good agreement with our
results. In the
atomistic  
simulations, performing the scaling analysis is very
computationally demanding and perhaps not even realizable. 
The scaling of
the breaking stress in un-symmetric junctions with different arm
length of the participating dislocations was found to be more
complicated and will not be discussed here. 
 
Before presenting our concluding remarks, we point out the features
that are missed in our model and the effect they may have on the
results discussed thus far. In our simulations, the dislocations do not acquire
a jog as they go past each other. While this does not affect the
junction structure, it will alter the breaking stress, since the
external stress should supply the energy required to create the jog.
However, the jog contribution to the breaking stress is only a small
fraction of the stress required to unzip the
junctions\cite{saada}. In order to examine the effect of
the stress cut-off distance, $r_c$, we have also carried out all the above
simulations by choosing $r_{c}$ to equal $b/2$ and $2b$. The junction
structure showed very little difference in the two calculations. Also,
the breaking stress in all the cases was within 10$\%$ of the values reported
here. 

In conclusion, we have developed an  efficient method
to study  dislocation interactions in FCC metals. We have
illustrated that the method can provide ``rules'' like critical
angle for junction formation and breaking stress criteria,
which can be used in 3D dislocation dynamics models. The results from
our simulations for junctions in different configurations, when
appropriately averaged, can provide parameters related to
junction strength in models for single crystal plasticity. Fits
to these parameters from
tension tests on single crystals have revealed a
hierarchy of junction strengths in FCC crystals\cite{franciosi}. Work is in progress
to verify the observed hierarchy on the basis of our simulations.
In the future we also hope to study other problems like the strength
of dislocation junctions in alloys and interactions of extended
dislocations with defects.

We are grateful to M. Fivel, A. Needleman, M. Ortiz and D. Rodney for
illuminating discussions.  We also appreciate the support of the
Caltech ASCI program and the Brown University MRSEC.

\vspace{-2cm}


\begin{references}
\bibitem{stack} The stacking fault energy of Al is 
obtained from the many-body potentials used in Ref.\cite{rodney},
while for Ag it is taken from Appendix 2 of Ref.\cite{hirthlothe}.
\end{references}
\end{document}